\documentclass[twocolumn,showpacs,preprintnumbers,amsmath,amssymb,floatfix,prstab]{revtex4}

\usepackage{graphicx}
\usepackage{subfigure}
\usepackage{dcolumn}
\usepackage{bm}
\usepackage{ctable}
\usepackage{multirow}

\begin{document}
\title{An Empirical Study of the Effect of Granting Multiple Tries for Online Homework}
\author{Gerd Kortemeyer}
\email{kortemey@msu.edu}
\affiliation{%
Lyman Briggs College and Department of Physics and Astronomy,
Michigan State University, East Lansing, MI 48824, USA
}%
\date{\today}
\begin{abstract}
When deploying online homework in physics courses, an important consideration is how many attempts should be allowed in order to solve numerical free-response problems. While on the one hand, the maximum number of
attempts should be large enough to allow learners mastery of the concepts and disincentivize copying, on the other hand, too many allowed attempts encourage counter-productive problem-solving behavior. In a study of an introductory calculus-based physics course for scientists and engineers which had different numbers of allowed attempts in different semesters, it was found that except for the initial tries, students are not making productive use of later attempts, and that in fact higher numbers of allowed attempts lead to less successful homework completion.
\end{abstract}

\maketitle

\section{Introduction}
Online systems have been found to be an effective and efficient means of giving homework with feedback to large enrollment courses~\cite{kashy2001,kortemeyer08}. Oftentimes, these courses do not employ a sufficient number of teaching assistants to provide graded homework in any other scalable way. Thus, the number of online homework systems is growing, for example Quest~\cite{utservice} (formerly University of Texas Homework Service), WeBWork~\cite{WebWork}, WebAssign~\cite{webassign}, Sapling~\cite{sapling}, Mastering~\cite{mastering}, as well as our LON-CAPA system~\cite{loncapa}, to name but a few.

A number of studies have been carried out regarding the educational effectiveness of these systems, at times with mixed results. Critics of these systems point out that the
electronic online medium can be in the way of employing higher order thinking skills. For example, Pascarella argued that  multiple possible attempts and instant feedback are 
``turning thinkers into guessers"~\cite{pascarella}. Indeed it was found that a large number of students admit to ``randomly guessing'' results~\cite{kortemeyer09}, and that typical times between subsequent submissions of tries to homework problems are in the range of a few seconds~\cite{kortemeyer09} --- too little time to actually think through the solution. Based on these results, the actual occurrence of random guessing may be in the range of 30\% of 50\% of all submissions, however, it has also been found that the amount of guessing can be somewhat reduced by introducing more frequent quizzes during classroom times~\cite{laverty2012}.

On the other hand, it was found that online homework is overall increasing student performance in physics~\cite{bonham2001,mestre2002,cheng2004} and elsewhere~\cite{richards2011}, although this is not necessarily attributable to the medium {\it per se}~\cite{bonham2003}. It is particularly helpful for learners on the brink of failing courses~\cite{kortemeyer08}, as well as for female learners, who take more advantage of the rich peer-to-peer interaction afforded by the problem randomization~\cite{kortemeyer09}.

Finally, there is the concern of copying homework, something which of course happens both with paper and pencil and online; however, this behavior is somewhat harder to detect in the online realm, as derivations are not submitted~\cite{palazzo10} --- having multiple attempts may  prevent students from desperately copying other students' work for fear of ``running out of tries'' of their own.

Most systems allow instructors to set the maximum number of allowed tries per homework problem. While for multiple-choice problem, there is in fact little room for variation (for example, it makes little sense to give 10 tries on a problem with 5 options), there is no such pre-determined limit for numerical free-response questions. A balance needs to be found between granting a sufficient number of attempts in order to allow learners mastery of a certain concept and disincentivizing copying versus encouraging random guessing or other counter-productive behavior. How does the maximum number of allowed attempts influence learner behavior?

\section{The Course}
\begin{figure}
\begin{center}
\includegraphics[width=0.45\textwidth]{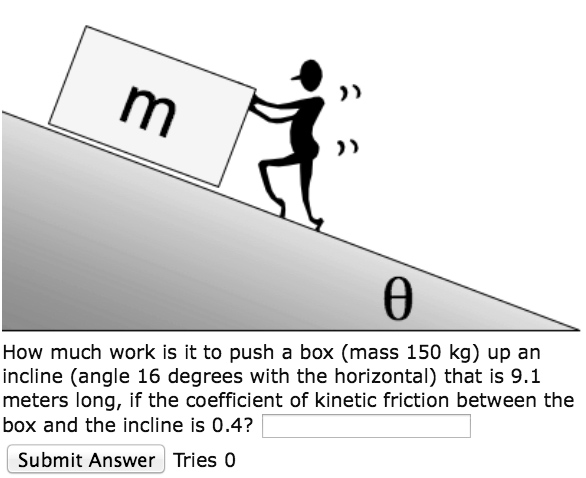}
\end{center}
\caption{Example of an online homework problem typical for the course under investigation. Numbers are randomizing between students so that simple copying of solutions is not possible.}
\label{fig:example}
\end{figure}

The course under investigation is the first semester of an introductory calculus-based physics course for scientists and engineers, dealing mostly with classical mechanics. This large enrollment course (typically having more than 500 students) is taught by different instructors in different semesters. While textbooks changed over the years, the overall structure of the course remained the same: more or less traditional lectures, several midterm exams, a final exam, and online homework. In all semesters, the vast majority of online homework were standard free-response numerical problems, expecting answers such as ``43 m/s'' or ``341.2 N'' (see Fig.~\ref{fig:example} as an example), i.e., problems that (as opposed to multiple choice) have ``infinitely'' many answer options. Immediate feedback on correctness was given, and no penalty was imposed for using multiple tries. In all semesters, over 200 online homework problems were assigned, leading to over 100,000 data points per course, where a data point is a particular student working on a particular problem --- each data point captures the number of tries used and whether or not the problem was correctly solved.

However, different instructors put slightly different overall weights on the online homework and, more notably, granted different numbers of maximum tries, see Table~\ref{tab:overview}. Some instructors chose to strongly limit the number of tries (10 or 12 tries), some were more lenient (20 tries), and one instructor decided to essentially put no limit on the number of tries (99 tries).

\section{Results}
\subsection{Overall Success and Failure on Homework}
As Table~\ref{tab:overview} shows, in all scenarios, students solved the vast majority of online homework. However, there is no clear correlation between the maximum number of tries granted and the relative weight on the one hand, and the percentage of problems correctly solved on the other. In fact, quite counter-intuitively, the percentage of correctly solved problems is the highest for the course with the lowest number of allowed tries.

When students are not arriving at the correct solution, there are essentially two scenarios: they give up on the problem before the tries are exhausted (i.e., abandon it), or they run out of tries. Not surprisingly, ``running out of tries'' is correlated with the maximum number of tries (and in the course with 99 tries, nobody runs out tries). 
\begin{table}
\begin{tabular}{cc|ccc}
Max.~Tries&Fraction Grade&Solved&Given Up&Out of Tries\\
\hline
10&20\%&98.0\%&1.6\%&0.4\%\\
12&25\%&96.2\%&3.4\%&0.4\%\\
20&20\%&97.0\%&2.9\%&0.1\%\\
99&20\%&91.8\%&8.2\%&0.0\%
\end{tabular}
\caption{\label{tab:overview}Overview of the online homework components of four semesters of the course under investigation. In different semesters, different numbers of maximum tries were granted, and one of the four semesters put more weight on the online homework. The table shows the percentage of homework solved, the percentage of problems that students gave up on, and the percentage of problems where the students ran out of tries.}
\end{table}
\subsection{Limited Tries Scenarios}
\begin{figure*}
\begin{center}
\includegraphics[width=0.48\textwidth]{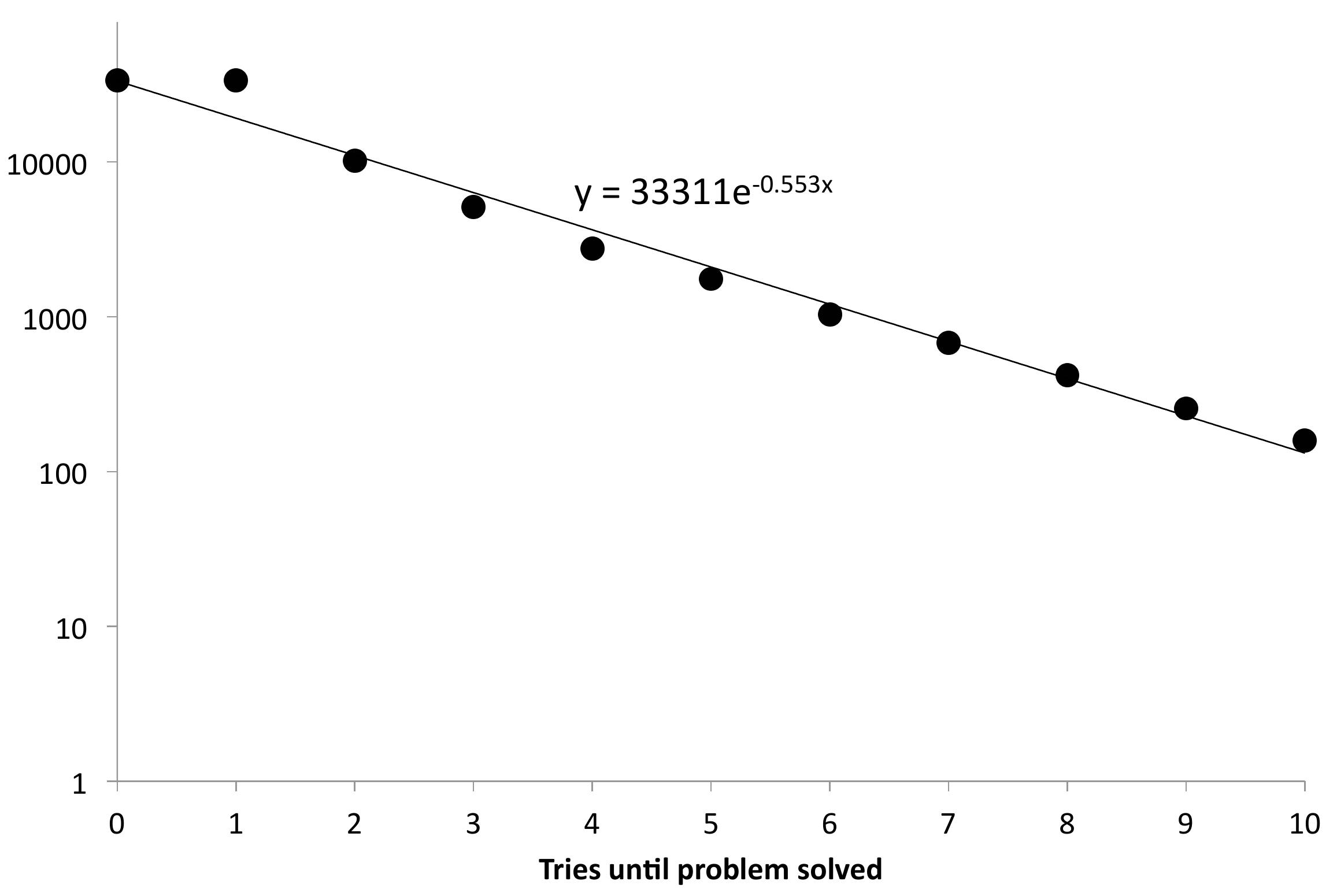}
\includegraphics[width=0.48\textwidth]{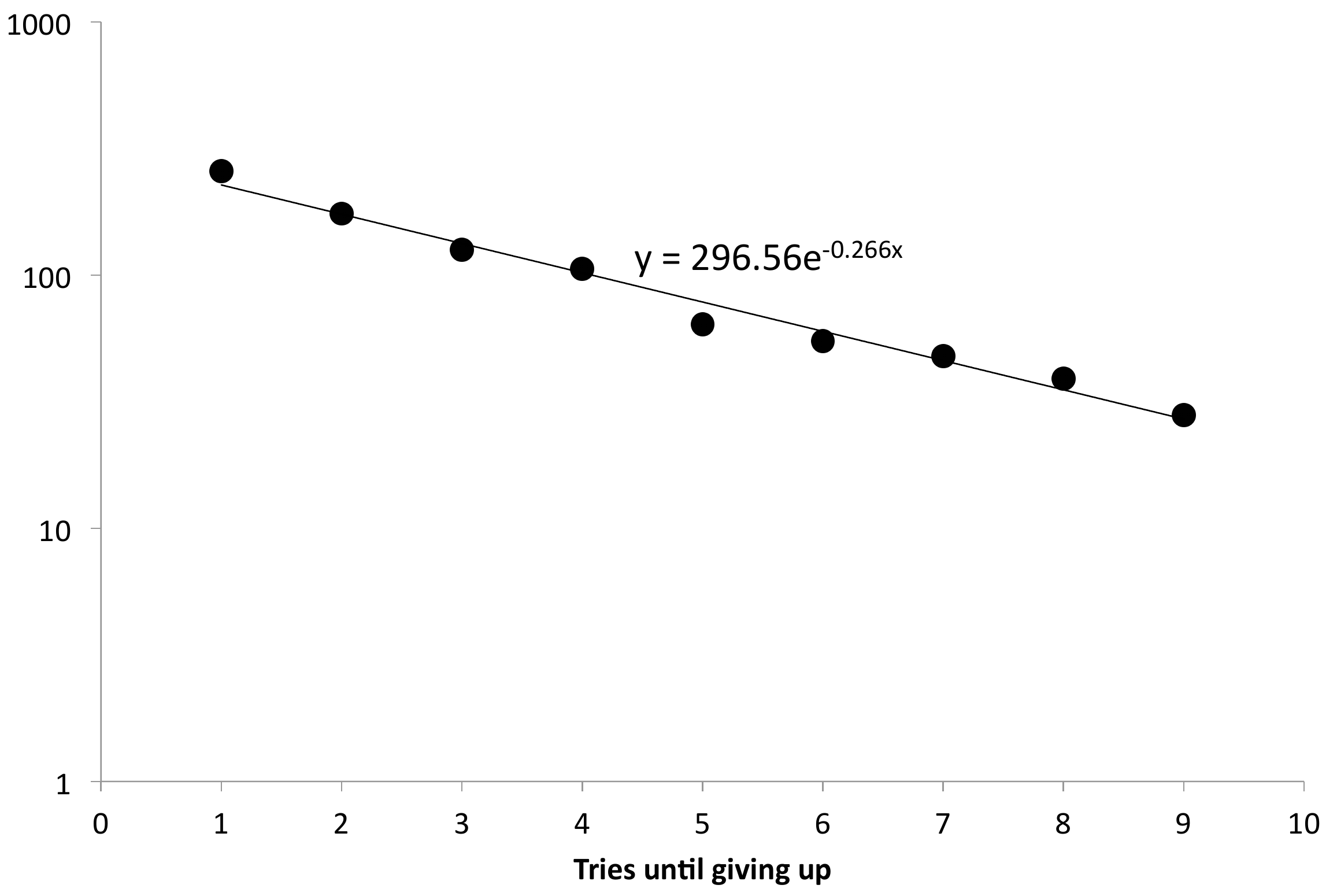}
\includegraphics[width=0.48\textwidth]{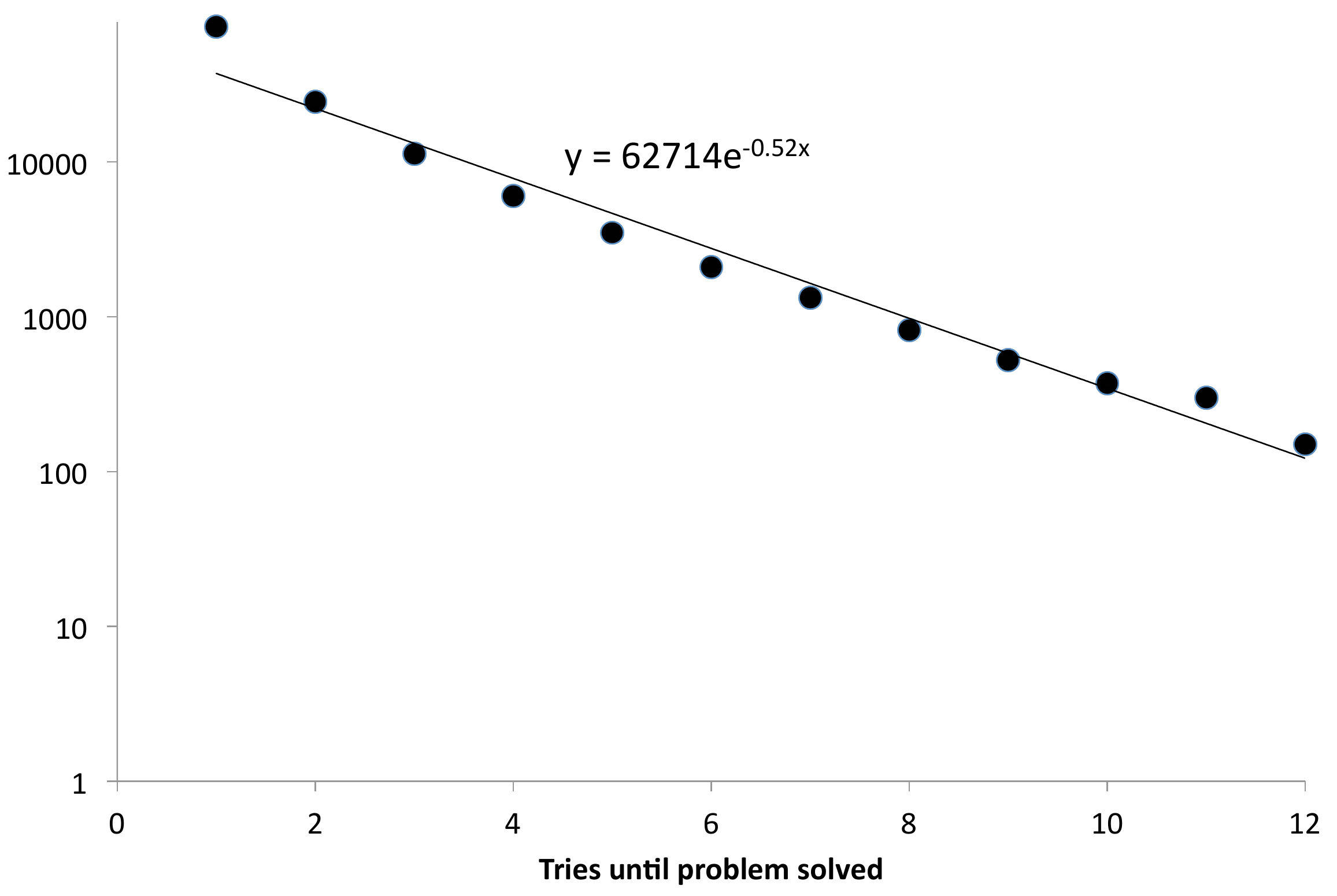}
\includegraphics[width=0.48\textwidth]{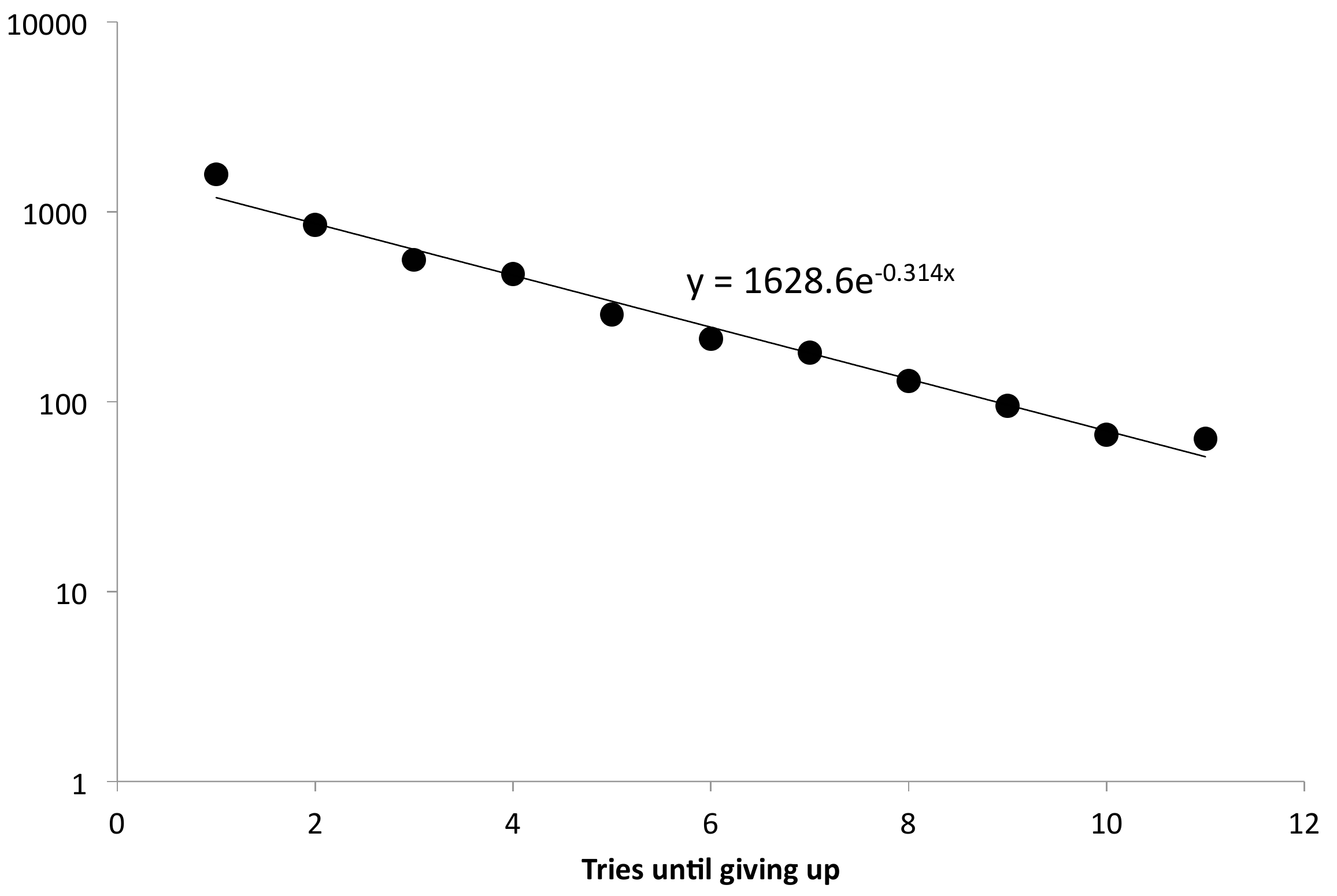}
\includegraphics[width=0.48\textwidth]{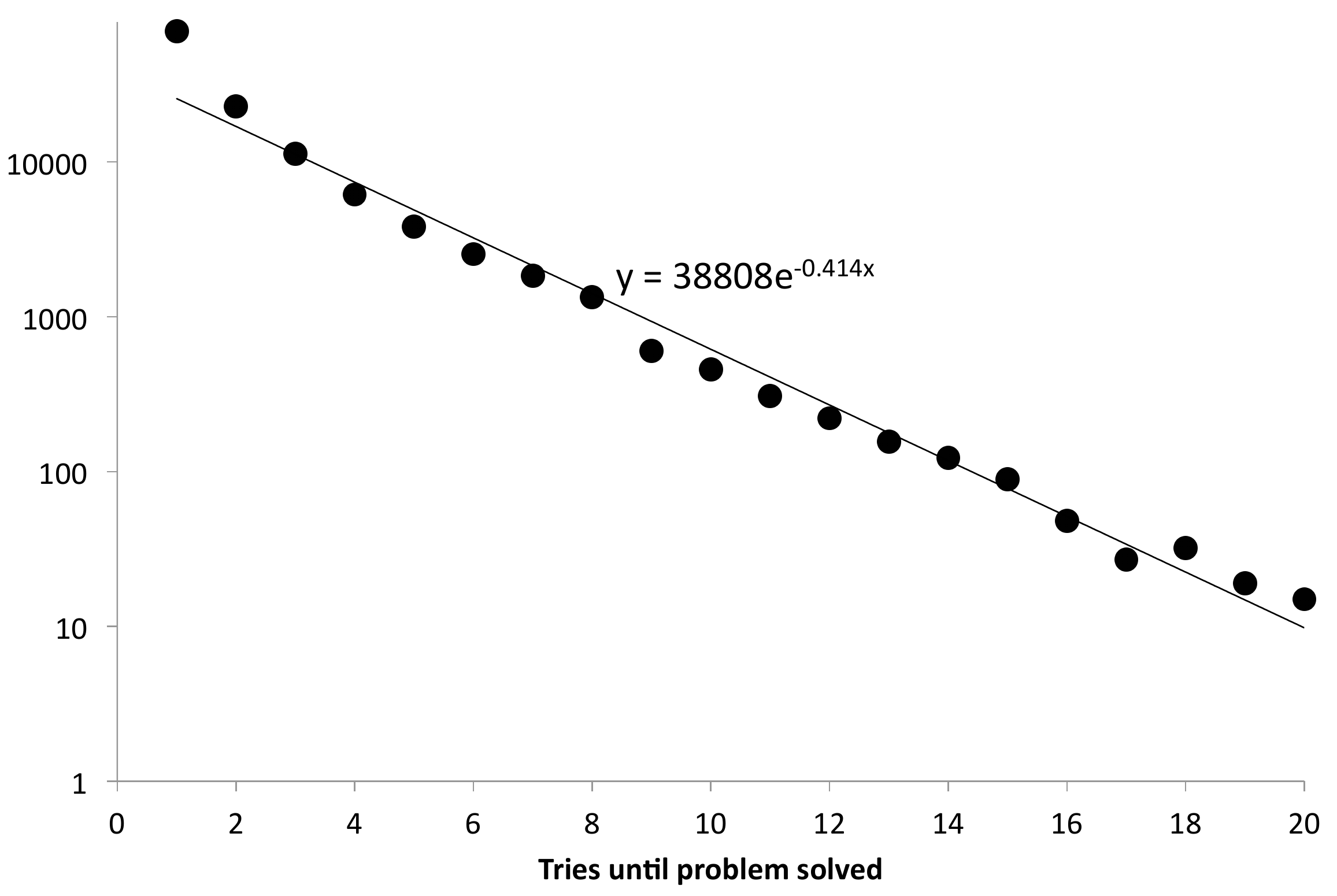}
\includegraphics[width=0.48\textwidth]{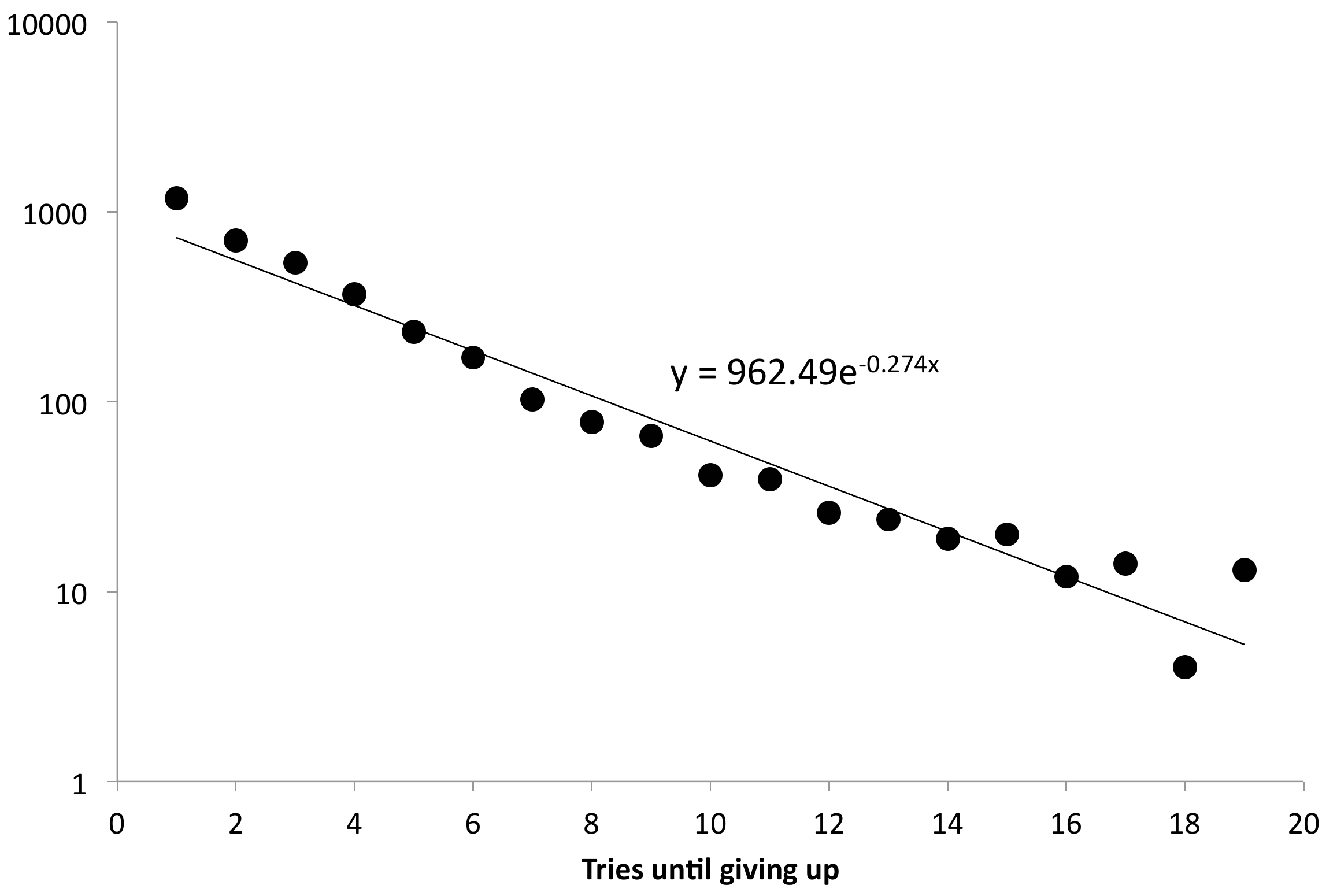}
\end{center}
\caption{Number of problems solved after the $n$-th attempt ($\Delta N_s(n)$, left panels) and given up on (abandoned) after the $n$-the attempt ($\Delta N_a(n)$, right panels) for different numbers of maximum allowed tries (10, 12, and 20 allowed tries, top to bottom, respectively). Indicated are also exponential fits to the data and the respective equations for the fits.}
\label{fig:limitedtries}
\end{figure*}

Looking into more detail of how students are taking advantage of multiple allowed tries is revealing: the left panels of Fig.~\ref{fig:limitedtries} show how many problems we were solved after the $n$-th attempt ($\Delta N_s(n)$), while the right panels show how many problems were given up on (abandoned) after the $n$-th attempt ($\Delta N_a(n)$). In other words, Fig.~\ref{fig:limitedtries} represents a histogram of the student/problem data points, sorted by outcome (solved or abandoned) and binned by the number of the last submitted attempt.

As the exponential fits in Fig.~\ref{fig:limitedtries} indicate, except for the first one or two attempts, the results are surprisingly compatible with
\begin{eqnarray}
\Delta N_s(n)&=&N_{s,0}\exp(-\lambda_s n)\label{eqn:success}\\
\Delta N_a(n)&=&N_{a,0}\exp(-\lambda_a n)\label{eqn:giveup}
\end{eqnarray}
where $N_{s,0}$ and $N_{a,0}$ are constants roughly representing the numbers of problems solved or abandoned during early attempts, and $\lambda_s$ and $\lambda_a$ determining how quickly the function falls off.  In other words, the number of problems solved or abandoned on a given try exponentially decays with the number of tries, and $\lambda_s$ and $\lambda_a$ are the respective ``decay constants.'' The absolute values of $N_{s,0}$ and $N_{a,0}$ are irrelevant, since they depend on how many students and problems the course in a given semester had (which varied), but their relative values within a given semester give an idea of the ratio of solved versus abandoned problems.

Interpreting these relationships is complicated by two facts: $n$ is not a continuous variable, and there are two ``decay channels'' drawing from the same pool of neither solved nor abandoned problems. Ignoring both complications, it is illustrative to write the above in differential form
\begin{eqnarray}
\frac{dN_s}{dn}&=&-\lambda_sN_{s,0}\exp(-\lambda_s n)=-\lambda_sN_s\label{eqn:successdiff}\\
\frac{dN_a}{dn}&=&-\lambda_aN_{a,0}\exp(-\lambda_x n)=-\lambda_aN_a\label{eqn:giveupdiff}
\end{eqnarray}
--- now $\lambda_s$ and $\lambda_a$ can be interpreted as {\it constant} probabilities of solving or abandoning a problem. Constant probabilities mean that the attempts are independent of each other, i.e., that the students did not learn from previous attempts, but are instead succeeding or throwing in the towel at constant rates between attempts. 

\begin{figure}
\begin{center}
\includegraphics[width=0.4\textwidth]{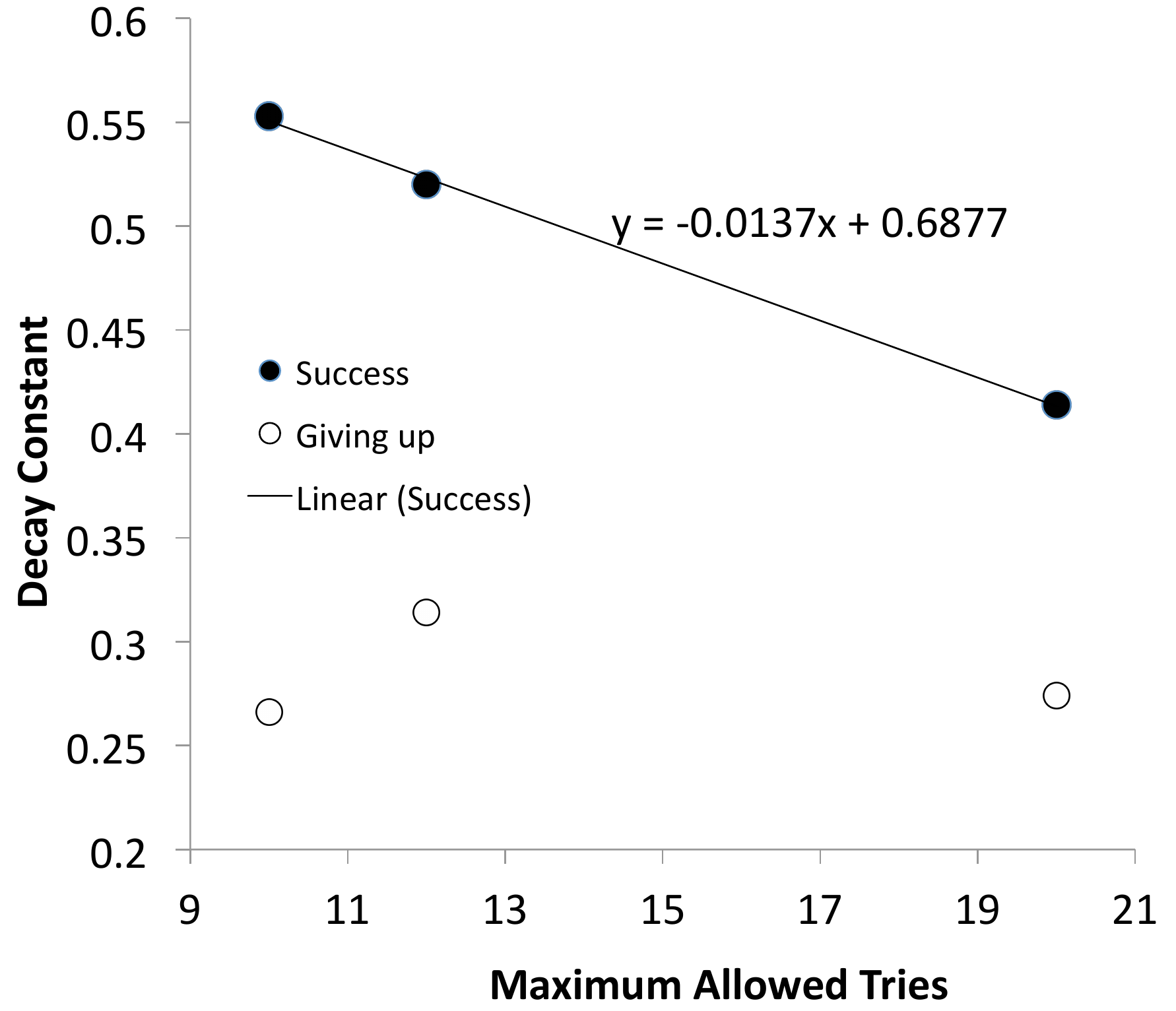}
\end{center}
\caption{``Decay constants'' $\lambda_s$ (solid dots) and $\lambda_a$ (open dots) for the two ``channels'' of online problem resolution as a function of maximum allowed tries. The decay constant $\lambda_s$ for solving the problem decreases roughly linearly, while $\lambda_a$ for ``giving up'' shows no such dependency.}
\label{fig:exponents}
\end{figure}

Fig.~\ref{fig:exponents} shows $\lambda_s$ and $\lambda_a$ as a function of the maximum number of allowed tries, extracted from the exponential fits. While the decay constants $\lambda_a$ for the ``giving up'' channel show no clear dependency on the number of allowed attempts (hovering between 0.25 and 0.35), the decay constants $\lambda_s$ for solving the problems decrease with increasing number of allowed attempts. As it turns out, the three $\lambda_s$ as a function of maximum tries happen to fit a linear relationship, which of course cannot be true over a wider range of tries. However, it is interesting to note that if one were to extrapolate this relationship to the limit of only one allowed attempt, it would end up with a probability of around 0.67, which is typical exam performance. At the other end, the linear relationship clearly has to break down, as it would indicate that for a maximum number of 50 tries, nobody would solve any problems anymore.

\subsection{The Unlimited Tries Scenario}
\begin{figure*}
\begin{center}
\includegraphics[width=0.48\textwidth]{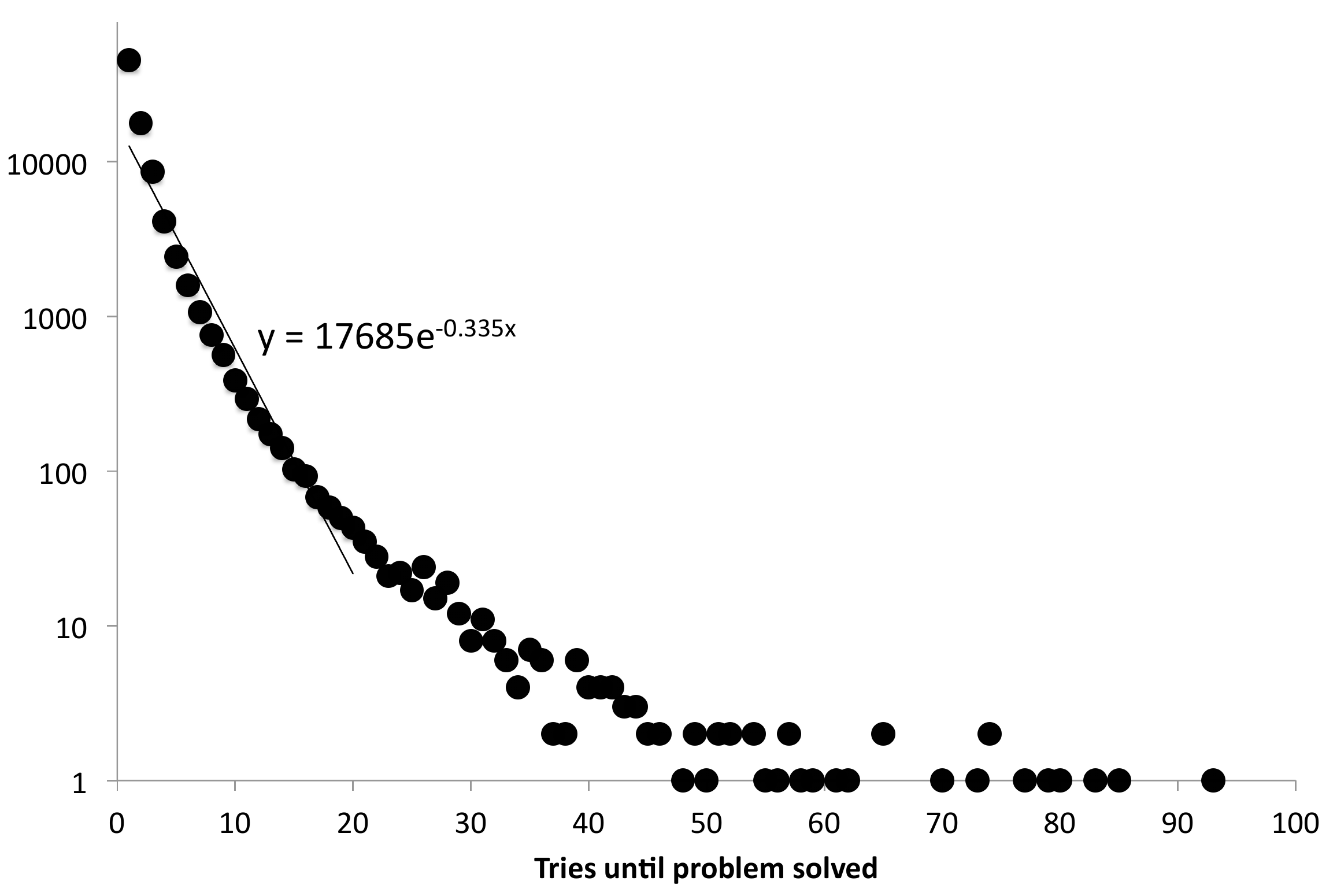}
\includegraphics[width=0.48\textwidth]{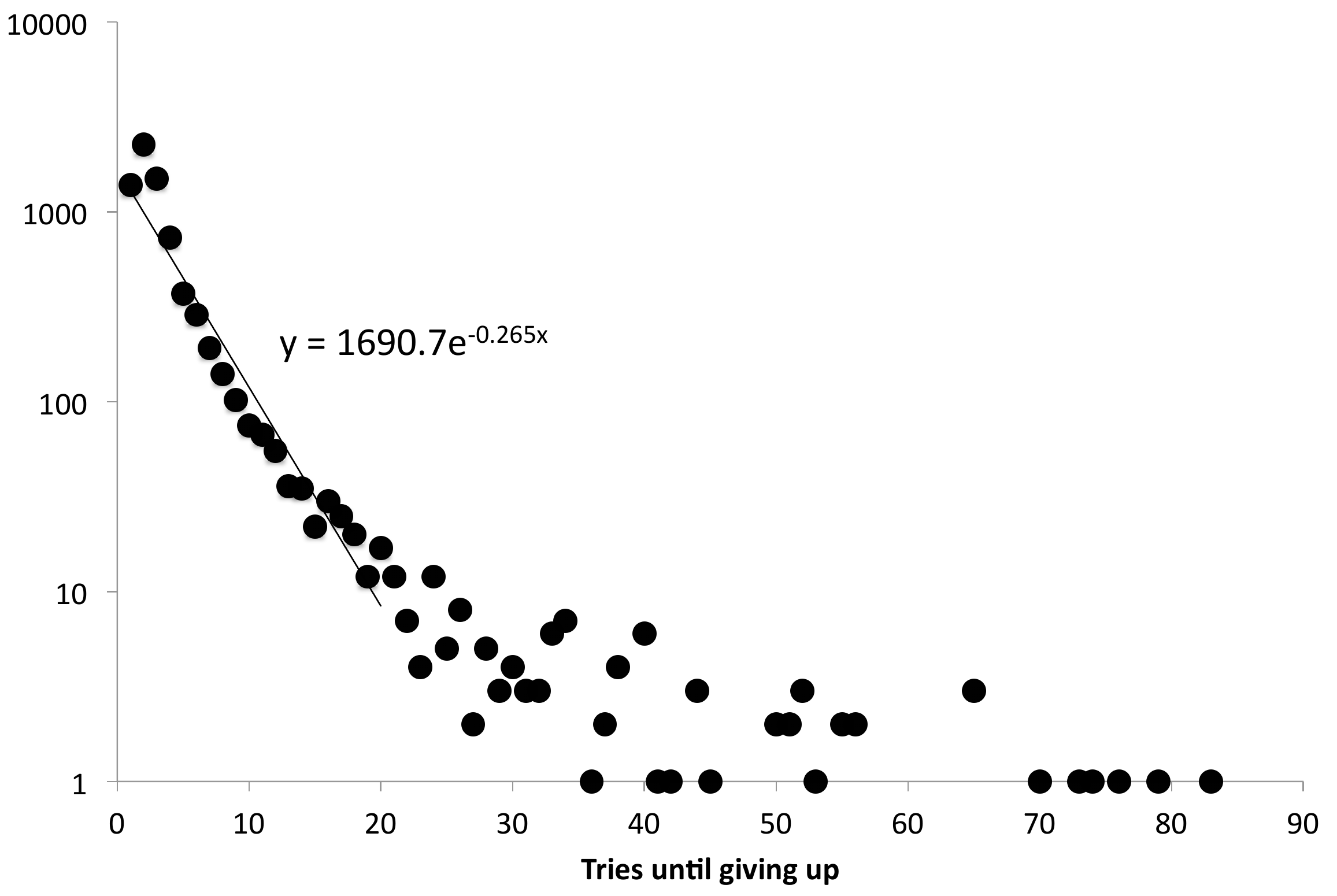}
\end{center}
\caption{Number of problems solved after the $n$-th attempt ($\Delta N_s(n)$, left panel) and given up on (abandoned) after the $n$-the attempt ($\Delta N_a(n)$, right panel) for a course with unlimited allowed tries. For comparison to Fig.~\ref{fig:limitedtries} and Fig.~\ref{fig:exponents}, an exponential fit to the first 20 attempts was performed for both scenarios.}
\label{fig:unlimitedtries}
\end{figure*}

While the exponential decay model is surprisingly robust for limited attempts, it of course eventually has to break down, and Fig.~\ref{fig:unlimitedtries} shows the same characteristics for the 99 (``unlimited'') tries scenario. While the dependencies start out exponential (as in the case of limited tries), eventually the histograms flatten out (after all, we are dealing with integer numbers of solved or abandoned problems).

For comparison, an exponential fit to just the first 20 attempts was performed. As expected, the decay constant $\lambda_a$ of 0.265 for giving up on the problems falls into the range of the results for limited tries (see Fig.~\ref{fig:exponents}), and the decay constant $\lambda_s$ of 0.335 for succeeding in solving the problem follows the trend of less and less successful problem solving per attempt with increasing number of allowed attempts. In other words, having unlimited tries has no effect on the rate of students giving up on problems (one would have hoped for less discouragement), and instead hurts the success rate per attempt (less productive behavior, probably because individual attempts are taken less seriously). As the number of  opportunities for giving up increases with number of allowed attempts, overall less homework gets solved (see Table~\ref{tab:overview}).

\subsection{Learner Self-Regulation}
In the previous subsections, it was found that overall higher numbers of allowed attempts are correlated with less productive behaviors: smaller success rates on particular tries, and higher total numbers of students giving up on problems. Learners appear to fall into the trap of ``turning thinkers into guessers''~\cite{pascarella}. But is this really true across the board, or are certain segments of the student population more susceptible, while others might employ self-regulation where external regulation is missing?
\begin{figure*}
\begin{center}
\includegraphics[width=0.48\textwidth]{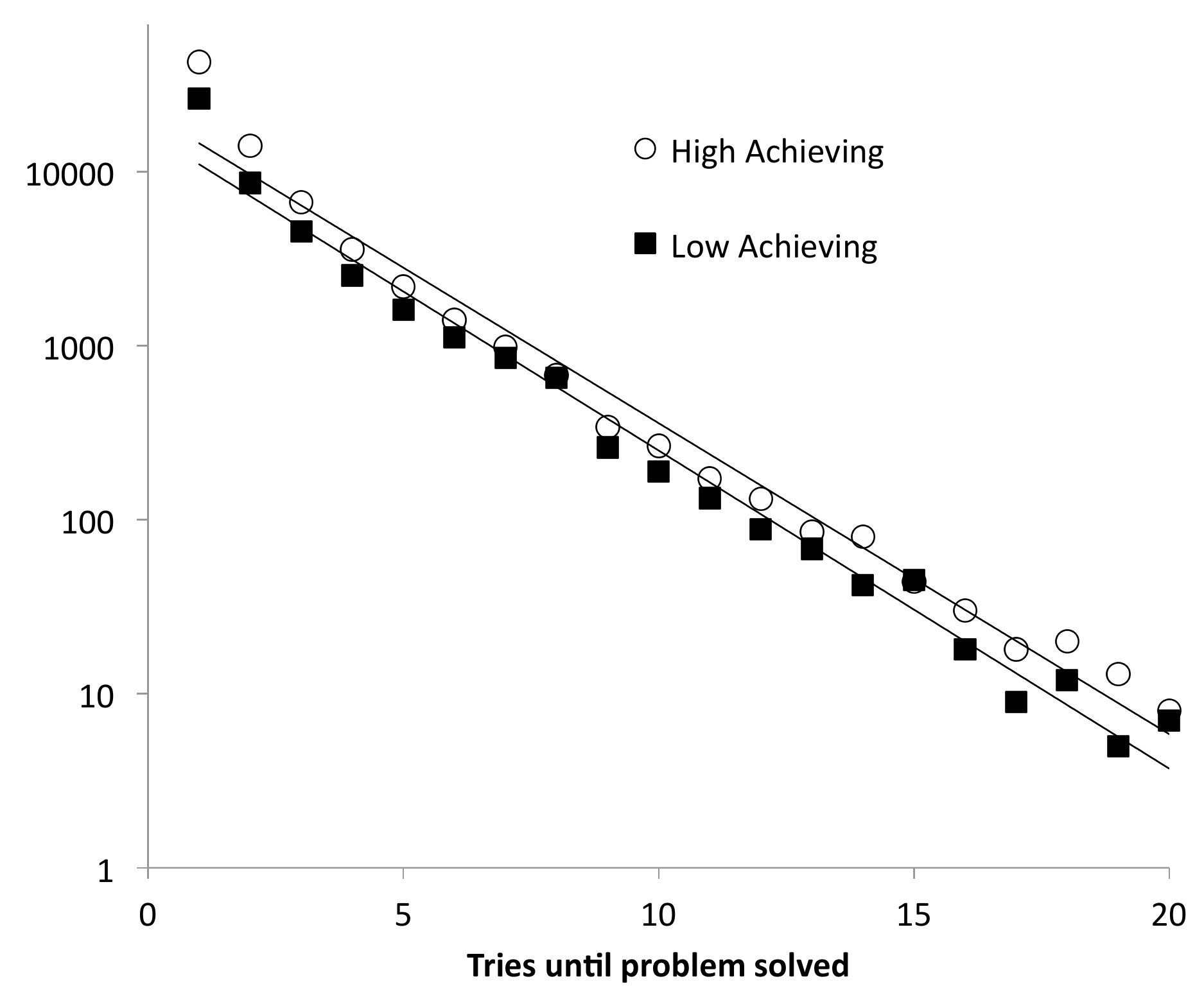}
\includegraphics[width=0.48\textwidth]{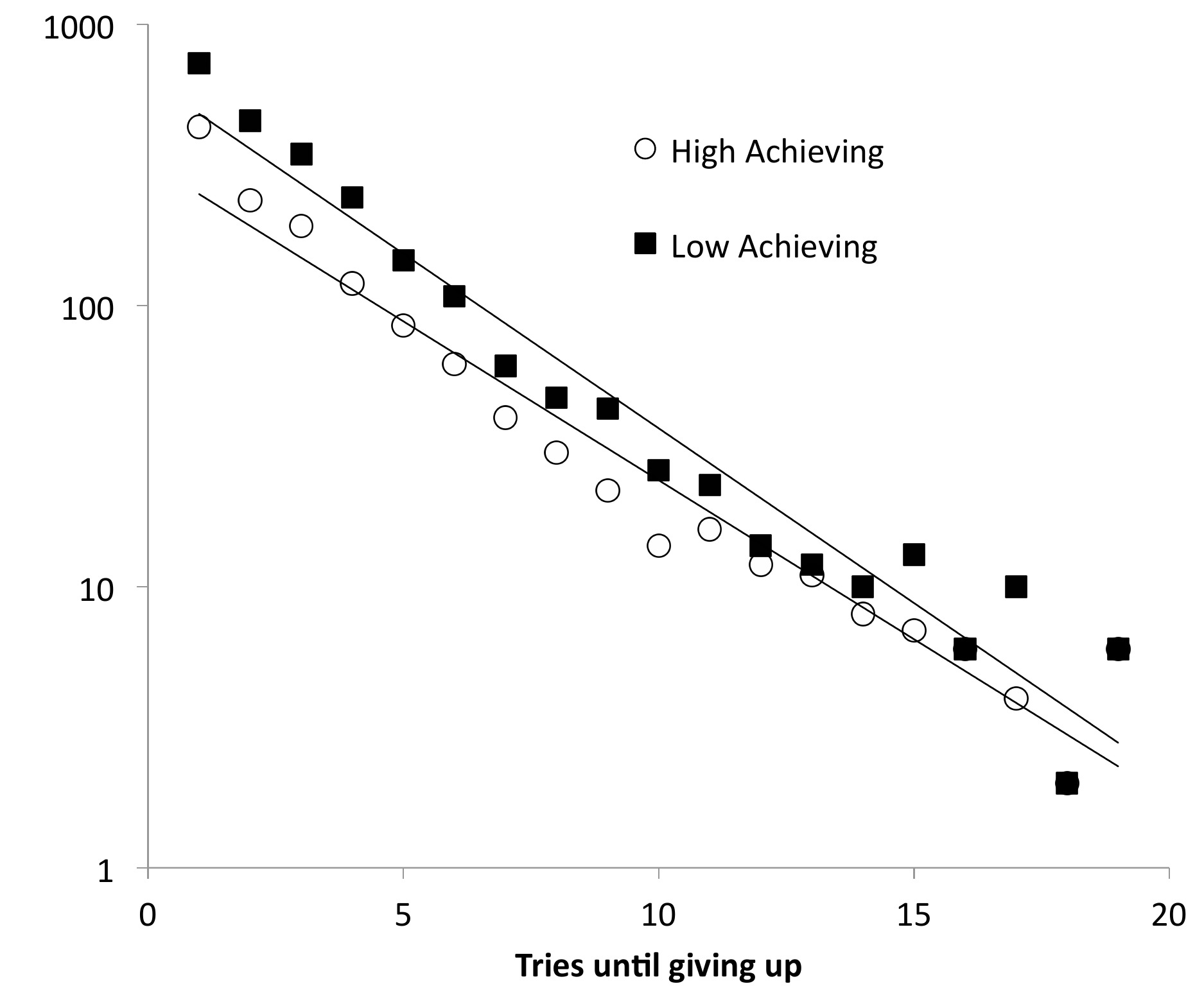}
\includegraphics[width=0.48\textwidth]{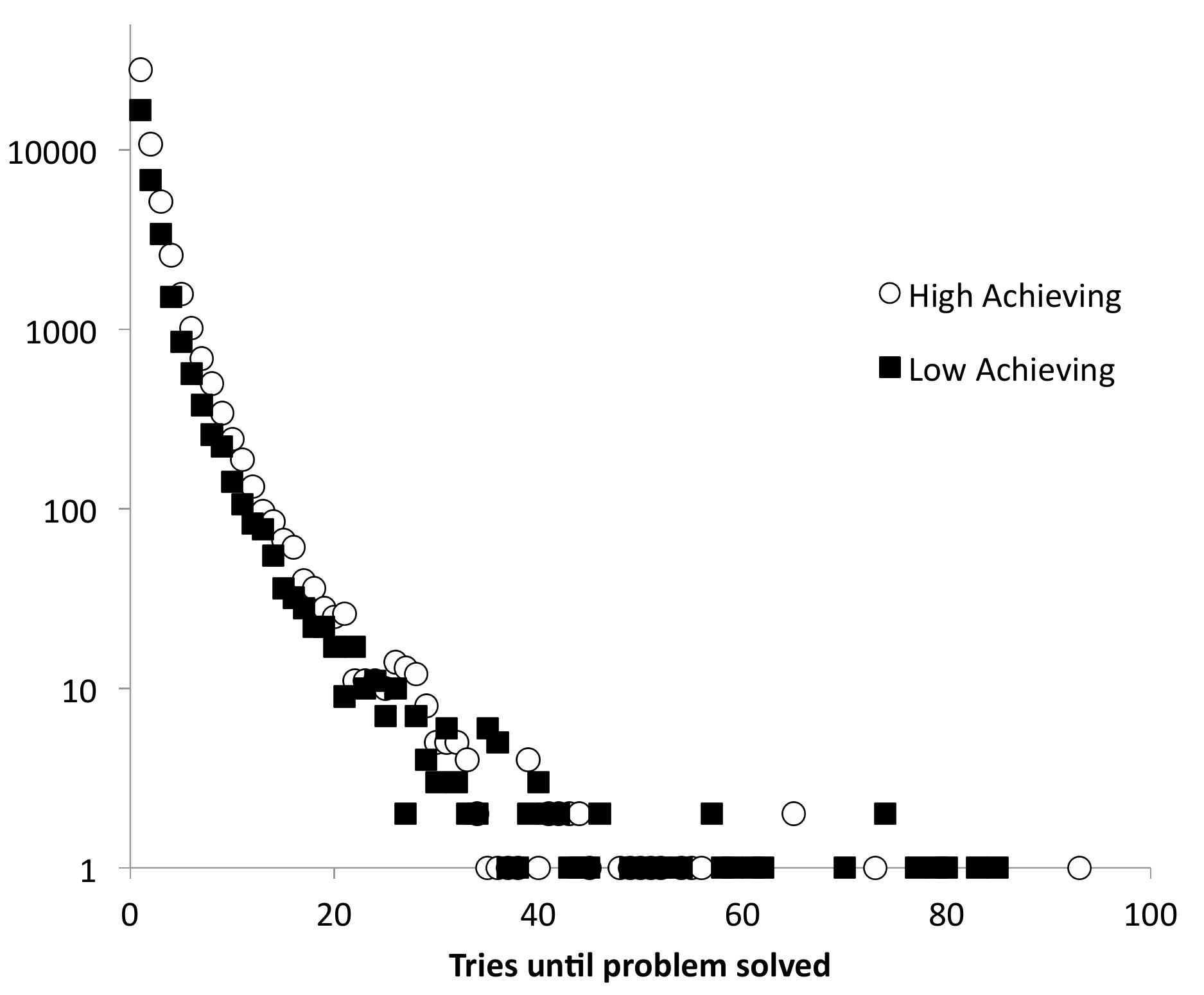}
\includegraphics[width=0.48\textwidth]{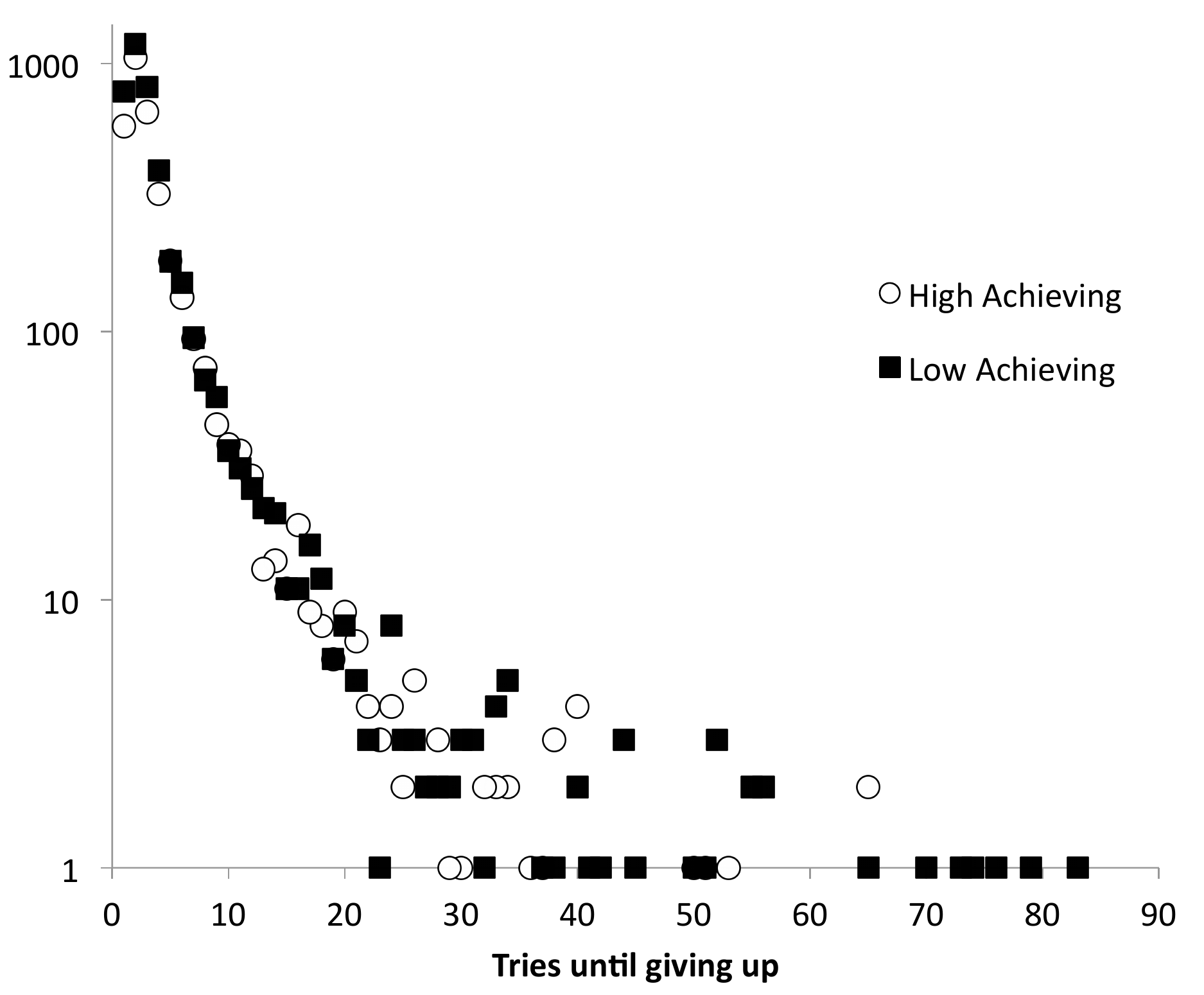}
\end{center}
\caption{Number of problems solved after the $n$-th attempt (left panels) and given up on after the $n$-the attempt (right panels) for different numbers of maximum allowed tries (20 (top) and 99 (bottom) allowed tries), separated by students achieving high and low exam grades.}
\label{fig:highlow}
\end{figure*}

To investigate if high and low achieving students are making different use of the multiple attempts, exam grades were taken into account. Performance on homework was separated between students performing higher or lower than the median of the combined exam scores. Fig.~\ref{fig:highlow} shows the result of this analysis for the semesters with 20 and 99 maximum allowed tries, respectively.

It is evident that students achieving higher exam grades succeed on more and give up on less homework problems, i.e., in Eqns.~\ref{eqn:success} and \ref{eqn:giveup}, the $N_{s,0}$ and $N_{a,0}$ are different for the two student groups. However, there is no discernible difference in the slopes of the exponential fits; after the initial tries, both groups have similar ``decay rates'' $\lambda_s$ and $\lambda_a$ for both success and giving up, i.e., there is no difference in how they are making use of later tries.

While exam scores are certainly a measure of success in a course, at this point, it should at least be remarked that they are not necessarily a measure of learning. As noted, in their homework performance, high and low achieving students mostly differ in how successful they are on the initial attempts --- on exam problems, students only have this one ``initial'' attempt, so it may not be surprising that this is how and where the groups differ.  In any case, even higher achieving learners appear to be turned into guessers rather than thinkers as they submit more and more tries.

\section{Discussion}
The most discouraging result of this study is that the probability for solving or giving up on a problem on a given attempt is constant for reasonably limited numbers of maximum tries. This means that the probability is independent of previous attempts; the students did not learn or in any other way profit from previous attempts. One would have hoped instead that the probability would increase on later attempts as students are gaining a deeper understanding of the problem. The only notable deviations from the ``exponential decay'' occur for the first or second attempt in limited tries scenarios (most likely due to ambitious students getting the problem correct early on, or due to copying of answers), as well as for the late attempts in unlimited tries scenarios (where the occasional problem or two get solved).

How can these constant probabilities be explained? It cannot just be completely random guessing, as that would not be successful at all on a problem with infinite answer options. There may be a constant background of ``guessing'' in the form of hunches, trying things out, or ``educated'' guesses. This strategy may also account for the fact that a higher number of allowed attempts results in less probability of success per attempt, as the students may be lured into becoming less  risk-averse in their ``guesses.''

But guessing alone would not explain the data. Apparently not learning from previous attempts can also be due to unsystematic work: if solutions are not worked out systematically and symbolically step-by-step, but instead worked out using a ``plug-and-chug'' strategy (successively plugging numbers into formulas), the student cannot retrace steps and needs to start over every time. One can observe such behavior in help rooms: using their calculators, students are working their way through formula after formula, get some result, submit it, get it wrong, and then start again from zero. The online format encourages such unsystematic problem solving, as the answer derivations are not graded; students may believe that they can ``get away'' with sloppiness, while in fact, they are creating extra work for themselves by not profiting from previous attempts.

The only good news about the constant probabilities is that they seem to indicate relatively little copying on later attempts. Copying of answers may be part of the explanation for success in early attempts, but later attempts are likely the students' own work.

Of all semesters, the one with essentially unlimited tries had the worst homework performance (see Table~\ref{tab:overview}): only about 92\% of the homework was eventually correctly solved. While (naturally) no student ran out of tries, overall more students simply gave up on the homework (having less chance of success per try and more opportunities to give up).  

Students who do well on exams and those who do not are making very similar use of later tries, suggesting that self-regulation plays a minor role compared to the externally imposed maximum number of tries. Both high and low achieving students get equally disadvantaged by offering too many tries.

The results naturally would look very different for multiple-choice questions, as they have a limited number of answer options --- as opposed to the unlimited options of free-response questions, one would expect to see the effects of exhausting options on later tries. If also those question types would be solved randomly, the scenario would be similar to typical ``drawing without replacement'' with a variable probability between tries. However, due to the overwhelming prevalence of numerical free-response homework, insufficient data was available to investigate the effect of multiple tries on multiple-choice problems.

The results put into question the usefulness of standard numerical free-response questions --- not learning from previous failures, even among high achieving students, indicates a lack of higher order metacognitive thinking in favor of trial-and-error.  Truly open-ended questions or carefully crafted multipart questions may yield more desirable results, and most online homework systems would be capable of administering them, but unfortunately also here sufficient data is currently lacking.

\section{Conclusion}
Based on the results from four semesters of a large enrollment introductory physics course for scientists and engineers, it becomes apparent that granting a large number of maximum allowed tries for free-response online homework is not beneficial.

The rate at which students are succeeding per try decreases with increasing number of allowed tries, and the rate at which students give up on problems per try does not depend on the number of allowed tries.

Rather than giving students a better chance of succeeding and mastering the concepts, having a large or even unlimited number of tries appears to lead to less and less desirable problem solving strategies: the students do not profit from their previous attempts. Giving high number of maximum tries may be motivated by the desire to not have the students run out of tries in their attempt to master a concept, but as it turns out, running out of tries is a far less likely reason for failure than simply giving up.

If free-response online homework is to be used in a course, giving the students less allowed attempts (10 seems reasonable) overall leads to better results and (relatively) more desirable student behavior. Putting no limits on the number of allowed tries is, based on these results, definitely a mistake.

\begin{acknowledgments}
The author would like to thank Anne Denton and Gary Westfall for useful discussions.
\end{acknowledgments}

\bibliographystyle{apsper}
\bibliography{bibfile}

\end{document}